\documentclass[aps, pra, amsfonts, nofootinbib, letterpaper, showpacs
	, twocolumn
     ]{revtex4-1}

\usepackage{amsmath}
\usepackage{amssymb}
\usepackage{bm}
\usepackage{graphicx}

\newcommand{\dotp}{\bm{\cdot}}
\newcommand{\mh}[1]{\mb{\hat{#1}} }
\newcommand{\Lm}{\Lambda}
\newcommand{\der}[2]{\frac{d#1}{d#2}}
\newcommand{\tx}[1]{\textrm{#1}}
\newcommand{\pder}[2]{\frac{\prt#1}{\prt#2}}
\newcommand{\del}{\nabla}
\newcommand{\om}{\omega}
\newcommand{\tht}{\theta}
\newcommand{\pdop}[1]{\frac{\prt}{\prt#1}}
\newcommand{\sg}{\sigma}
\newcommand{\ul}[1]{\underline{#1}}
\newcommand{\bt}{\beta}
\newcommand{\Om}{\Omega}
\newcommand{\dl}{\delta}
\newcommand{\al}{\alpha}
\newcommand{\eps}{\epsilon}
\newcommand{\lm}{\lambda}
\newcommand{\mb}[1]{\mathbf{#1}}
\newcommand{\abs}[1]{\left|#1\right|}
\newcommand{\prt}{\partial}
\newcommand{\Gm}{\Gamma}
\newcommand{\tand}{\textrm{ and }}
\newcommand{\lr}[1]{\left(#1\right)}
\newcommand{\mf}[1]{\mathfrak{#1}}
\newcommand{\evals}[1]{\left.#1\right|}

\newcommand{\Sg}{\Sigma}

\begin{document}


\title{Laser-assisted electron-argon scattering at small angles}
\author{Nathan \surname{Morrison}}
\author{Chris H. \surname{Greene}}
\altaffiliation{Current address as of August 2012,
 Physics Dept., Purdue University, West Lafayette, IN 47907}
\affiliation{Department of Physics and JILA, University of Colorado, 
                  Boulder, Colorado 80309-0440, USA}

\begin{abstract}
Electron-argon scattering
 in the presence of a linearly polarized, low frequency laser field
 is studied theoretically.
The scattering geometries of interest are small angles
 where momentum transfer is nearly perpendicular to the field,
 which is where the Kroll-Watson approximation
 has the potential to break down.
The Floquet $R$ matrix method
 solves the velocity gauge Schr\"odinger equation,
 using a larger reaction volume than previous treatments
 in order to carefully assess the importance
 of the long range polarization potential
 to the cross section.
A comparison of the cross sections calculated
 with the target potential fully included
 inside 20 and 100 a.u.~shows no appreciable differences,
 which demonstrates that the long range interaction can not account for
 the high cross sections measured in experiments.
\end{abstract}

\maketitle

\section{Introduction} \label{sc_Introduction}

Since the development of intense lasers,
 investigated the modification of
 familiar processes by the presence of coherent light.
Particles that collide in the presence of a laser field
 can also exchange energy with the field
 in multiples of the laser frequency,
 a process known as laser-assisted collision.
This paper investigates free-free transitions
 in electron scattering,
 where the state of the target atom remains unaltered.

In 1973 Kroll and Watson derived an expression
 for the cross section of a laser assisted scattering event
 in terms of the elastic field-free cross section.
Their main assumptions were that the time duration
 of the interaction
 is short compared to the laser cycle,
 and that the target itself is unperturbed by the laser.
They found the following expression, \cite{KrollWatson1973}
\begin{equation}
\der{\sg_\nu}{\Om} \lr{\mb k_f,  \mb k_i}
			= \frac{k_f}{k_i}
				J_\nu^2(x) \der{\sg_\tx{el}}{\Om} \lr{\eps,  \mb Q}
\end{equation}
where $\nu$ denotes the number of laser photons absorbed
 by the electron
 ($-\nu$ is the number of photons emitted),
 $\mb k_f$ and $\mb k_i$ are the momenta
 of the final and initial electron,
 $J_\nu$ is the Bessel function of the first kind of order $\nu$,
 and $\der{\sg_{\mathrm{el}} } \Om$
 is the field free elastic cross section.
With the time dependent laser vector potential
 of the form $\mb A(t) = \mb A_0 \sin(\om t)$,
 the other parameters are defined as follows:
\begin{align}
 \mb Q &= \mb k_f - \mb k_i
 \\
x &= - \frac{e \mb A_0 \dotp \mb Q}{ m c \om }
 \\
\eps &= \frac{k_i^2}{2m}
- \nu \om \frac{\mh A_0 \dotp \mb k_i}{\mh A_0 \dotp \mb Q}
	 + \frac{m \lr{\nu \om}^2}{2 \lr{\mh A_0 \dotp \mb Q}^2}
\end{align}
In addition to assuming that the laser frequency $\om$
 is small,
 the Kroll-Watson approximation (KWA) assumes
 that the dimensionless quantity
\begin{equation} \label{AdotQ}
 \frac{e}{m c \om} \mb A_0 \dotp \mb Q
\end{equation}
 is sufficiently large that only on-shell scattering contributes.%
\footnote{See section 5 of \cite{KrollWatson1973}}
This assumption becomes questionable not just when
 the parameters of the laser are changed,
 but also in certain critical scattering geometries
 where $\mb A_0 \dotp \mb Q \approx 0$.

The group of Wallbank and Holmes
 has performed experiments investigating these geometries
 for several neutral targets,
 beginning with helium and argon
 \cite{Wallbank1993, Wallbank1994, Wallbank1994Ar}.
Their results show cross sections orders of magnitude
 greater than the Kroll-Watson prediction in regions where
 $\frac{e}{m c \om} \mb A_0 \dotp \mb Q$
 is small.
One of the first ideas proposed to explain this discrepancy
 was the polarization of the target by the field of the laser,
 however several separate theoretical treatments
 \cite{Rabadan1994, Geltman1995, Varro1995}
 showed such an effect to be negligible.
It was also shown that for certain densities of the target gas,
 double scattering could account for the experimental signal
 \cite{Dickinson1996dbl}.
Their determination of the experimental density was approximate,
 however, and to our knowledge the density dependence
 has not been confirmed experimentally.

\citet{Madsen1995} explore the derivation of
 the KWA,
 and they develop a generalized approximation
 by expanding the $T$ matrix for weak fields and soft photons,
 but without assuming (\ref{AdotQ}) is large.
The region where the KWA loses its validity is therefore avoided.
Their calculations show a few scattering geometries
 where the differential cross section is comparable to the experiments.
The shape of the experimental cross sections disagrees
 with the Ref.~\cite{Madsen1995} theory, though,
 and there would have to be a very large uncertainty in the determination
 of the elecron scattering angle
 for their calculations to explain the experimental cross sections
 at all angles.

The above treatments assume, as does the KWA,
 that only on-shell terms, i.e. terms where energy is conserved,
 contribute to the scattering event.
A few later works use approximations that include off-shell contributions.
\citet{Sun1998} apply the second Born approximation.
\citet{Jaron1997, Jaron1999}
 also use a similar off-shell approximation. 
They suggest that a diffraction effect due to a long range potential,
 i.e. an interaction with extent large compared to the electron
 deBroglie wavelength,
 could give rise to the sorts of cross sections at small angles
 seen in the experiment.
The results of these in comparison with our calculations
 are discussed further in section \ref{sc_Results}.

The advantage of $R$ matrix methods
 is that they provide an exact solution of the Schr\"odinger equation
 within the chosen reaction volume,
 and so are limited only by the size of that volume
 and the numerical methods used in the calculation.
\citet{ChenRobicheaux96} use a mixed gauge $R$ matrix method
 with a reaction volume of 30 a.u. 
They calculated cross sections with similar order of magnitude to the KWA.

The goal of this work is to use an exact $R$ matrix method
 with an expanded reaction volume.
Including a longer range for the electron to interact with the
 induced dipole potential of the argon atom
 will clarify what contribution the long range interaction
 has to the laser assisted cross section.
In section \ref{sc_Floquet_R} the details of the Floquet $R$ matrix method
 in the velocity gauge
 are laid out,
 and in section \ref{sc_Scattering_laser} the connection to scattering states
 and the form of the scattering matrix are derived.
Section \ref{sc_Results} discusses our numerical results,
 and section \ref{sc_Conclusions} summarizes our findings.
Atomic units are used throughout the rest of this paper.

\section{Floquet $R$ matrix method} \label{sc_Floquet_R}
The time and angle dependence of the electron wavefunction
 is represented by expanding in a product set
 of spherical harmonic and Floquet basis functions.
\begin{equation}
\Psi^\bt(\mb r, t)
	 = \sum_{\nu,l}
	  \frac{F^\bt_{\nu l}(r)}{r} Y_{l0}(\Om)
	  \frac{e^{-i(E + \nu\om)t}}{\sqrt{2\pi/\om}}
\end{equation}
Here $\bt$ enumerates a complete set of linearly independent
 solutions of the Schr\"odinger equation.
In order to make the treatment of larger reaction volumes tractable,
 this paper treats only scattering geometries
 where the incoming electron is parallel to the laser polarization.
The cylindrical symmetry of this case
 allows setting $m_z = 0$ for the entire calculation,
 considerably reducing the number of basis functions needed.

Taking $\bt$ as a column index,
 and combining the others into a row index,
 the radial functions can be thought of as comprising a matrix $\ul F(r)$.
With this the logarithmic derivative, or $R$ matrix, is defined as
\begin{equation}
\ul R (r)
 = \ul F (r) \lr{\ul F'(r)}^{-1}
\end{equation}
The $R$ matrix describes the behavior of the channel functions
 at the surface of a volume of constant radius $r$.

The $R$ matrix is found by solving the Hamiltonian
 in the velocity gauge for an electron scattering off of a potential.
The wavelength of the laser is around $2\times 10^5$ a.u.,
 which is much larger than the region of interaction of interest,
 so the vector potential is essentially constant in space.
This leads to the following form for the Hamiltonian:
\begin{equation}
 H
 = \frac{\mb p^2}{2} - \frac{1}{ c} \mb A \dotp \mb p
  + V(r)
\end{equation}
The target atom is represented by a model potential,
 borrowed from Chen and Robicheaux,
\cite{ChenRobicheaux96}
 containing a shielded Coulombic core
 and a long range induced-dipole term.
\begin{equation}
V(r) = -\frac Zr e^{- a_1 r}
					- a_2 e^{-a_3 r}
					- \frac{\al}{2 r^4} \lr{1-e^{-(r/r_{\mathrm{cut}} )^3}}^2
\end{equation}
Here $Z = 18$, the atomic number,
 and $\al = 10.77$, the argon polarizability.
The other parameters of the model potential,
 which were fitted to the field free phaseshifts for argon,
 are $a_1 = 3.04$,
 $a_2 = 10.62$,
 $a_3 = 1.83$,
 $r_{\mathrm{cut}} = 1.76$.

\subsection{Variational principle for the $R$ matrix}

An extension of the eigenchannel $R$ matrix method,
 adapted for the Floquet formalism and for the velocity gauge,
 yields the solution to the Schr\"odinger equation.
The solution is calculated numerically within a finite reaction volume $\Sg$.
It is helpful to begin with the Schr\"odinger equation
 for the velocity gauge hamiltonian in integral form.
(Note that, for notational brevity,
  we employ notation commonly used in differential geometry
  throughout this section.
 Function arguments and differentials are omitted from the integrands,
  but the integral is unambiguous
  as the domain of integration is denoted as a subscript on the integral sign.)
\begin{equation}
 \int_{\Sg,T} \Psi^* i \pder \Psi t
 = \int_{\Sg,T} \Psi^* \lr{
 	-\frac 12 \del^2 \Psi
	- \frac{i} c \mb A \dotp \del \Psi
	+ V(r) \Psi
    }
\end{equation}
The full version of the energy operator is necessary
 due to the fact that, in the Floquet formalism,
 wavefunctions can be superpositions of different pseudoenergy states.
The only restriction on the space of wavefunctions considered
 is that the spatial inner product between any two wavefunctions
 is periodic,
 i.e. 
\begin{equation} \label{periodic_assumption}
 \int_\Sg \Psi_1^* \Psi_2(t) = \int_\Sg \Psi_1^* \Psi_2 (t + T)
\end{equation}
This is a reasonable assumption based on the fact
 that the Hamiltonian is itself periodic.

Application of the first Green identity to the kinetic term
 gives a term with the derivative on the surface $\prt \Sg$.
\begin{align} 
 \frac 12 \int_{\prt \Sg, T} \Psi^* \pder \Psi n
 = \int_{\Sg,T}
    \Bigg(
 	\frac 12 \del \Psi^* \dotp \del \Psi
	\hspace{40pt}\\ \nonumber
	- \frac{i} c \mb A \dotp \Psi^* \del \Psi
	+ V(r) \Psi^* \Psi
   - \Psi^* i \pder \Psi t
    \Bigg)
\end{align}
This is the usual starting point to find a variational principle
 for the logarithmic derivative.
Because the velocity gauge contains a first derivative term, however,
 this must be taken into account in order to construct
 a variational principle.
Using a form of the divergence theorem, the identity becomes
\begin{equation}
 \begin{matrix} \displaystyle 
 \frac 12 \int_{\prt \Sg, T} \lr{
    \Psi^* \pder \Psi n
	+ \frac ic \mb A \dotp \mh n
	 \Psi^* \Psi
 }
 \hspace{\fill} \\ \displaystyle \hspace{10pt}
 = \int_{\Sg,T}
   \Big(
 	\frac 12 \del \Psi^* \dotp \del \Psi
	- \frac{i}{2c} \mb A \dotp 
	 \lr{ \Psi^* \del \Psi -  \del\Psi^* \Psi }
	\\ \hspace{\fill}
	+ V(r) \Psi^* \Psi
   - \Psi^* i \pder \Psi t
   \Big)
 \end{matrix}
\end{equation}

Now define the operator $\tilde L$ as follows:
\begin{equation} \label{tildeLdef}
 \tilde L
 = \pdop{n} + \frac i c \mb A \dotp \mh n \Psi
\end{equation}
The set of channel functions in all surface coordinates
 forms a linear space on
 the surface of $\Sg$.
Consider wavefunctions $\Psi_\bt$ that are eigenfunctions of $\tilde L$
 on this surface,
 i.e. $\tilde L \evals{\Psi_\bt}_{\prt \Sg} = b_\bt \evals{\Psi_\bt}_{\prt \Sg}$ .
The value $b_\bt$ is the usual logarithmic derivative with one term added.
\begin{equation}
b_\bt 
 = \lr{ \frac 1{\Psi_\bt} \pder{\Psi_\bt}{n}
    + \frac i {c} \mb A \dotp \mh n }_{\prt \Sg}
 = \lr{ \pder{\ln(\Psi_\bt)}{n} + \frac i {c} \mb A \dotp \mh n }_{\prt \Sg}
\end{equation}
The following functional is the variational principle
 for the eigenvalues $b_\bt$
 of the generalized logarithmic derivative operator:
\begin{equation} \label{varprinciple}
 b[\Psi]
 = 2
 \frac{ 
  \int_{\Sg,T}
   \lr{ \begin{matrix}
 	\frac 12 \del \Psi^* \dotp \del \Psi
	- \frac{i}{2c} \mb A \dotp 
	 \lr{ \Psi^* \del \Psi -  \del\Psi^* \Psi }
	 \\
	+ V(r) \Psi^* \Psi
   - \Psi^* i \pder \Psi t
    \end{matrix} }
  }
 {\int_{\prt \Sg,T} \Psi^* \Psi}
\end{equation}
It can be shown that the first variation
 $\dl b[\Psi_\bt]$ vanishes for all deviations $\dl\Psi$
 from the exact solution.
To show this, integrate by parts
 and use the periodicity restriction (\ref{periodic_assumption})
 on the energy operator.
\begin{equation}
 \dl \lr{\int_{\Sg, T} \Psi^* i \pder \Psi t}
 = \int_{\Sg, T} \dl\Psi^* i \pder \Psi t
  - \int_{\Sg, T} i \pder{\Psi^*}t \dl \Psi
\end{equation}
Using this,
 the variation can be written
\begin{widetext}
\begin{equation}
\begin{matrix}
 \dl b[\Psi_\bt] & \propto &
 \int_{\prt \Sg,T} \Psi_\bt^* \Psi_\bt
  \int_{\Sg,T}
   \lr{
 	\frac 12 \del \dl\Psi_\bt^* \dotp \del \Psi_\bt
	- \frac{i}{2c} \mb A \dotp 
	 \lr{ \dl\Psi_\bt^* \del \Psi_\bt -  \del \dl\Psi_\bt^* \Psi_\bt }
	+ V(r) \dl\Psi_\bt^* \Psi_\bt
   - \dl\Psi_\bt^* i \pder {\Psi_\bt} t
    }
 \\ & &
  -
  \int_{\Sg,T}
   \lr{
 	\frac 12 \del \Psi_\bt^* \dotp \del \Psi_\bt
	- \frac{i}{2c} \mb A \dotp 
	 \lr{ \Psi_\bt^* \del \Psi_\bt -  \del\Psi_\bt^* \Psi_\bt }
	+ V(r) \Psi_\bt^* \Psi_\bt
   - \Psi_\bt^* i \pder {\Psi_\bt} t
    }
  \int_{\prt \Sg,T} \dl\Psi_\bt^* \Psi_\bt
 + \mathrm{c.c.}
\end{matrix}
\end{equation}
\end{widetext}
The first Green identity can now be applied to each of the kinetic terms,
 and the divergence theorem can be applied to the vector potential terms,
 to show the variation vanishes.

\subsection{Solving for the $R$ matrix}

The numerical solution of the $R$ matrix is carried out
 by expanding the wavefunction in a basis set:
 $\Psi
 = \sum_{pi} c_{pi} \psi_{pi} (\mb r, t)
 = \sum_{pi} c_{pi} \frac{u_{p}(r)}{r} \Phi_i (\Om, t)
 $,
The radial basis functions $u_{p}(r)$
 can in principle be arbitrary.
In this calculation we have chosen a finite element DVR
 basis, of the kind described in \cite{Tolstikhin1996}.
The channel functions $\Phi_i (\Om, t) $ have the form
 $\Phi_i = 
	  Y_{l_i 0}(\Om)
	  \frac{e^{-i(E + \nu_i \om)t}}{\sqrt{2\pi/\om}}
 $.
The variational principle (\ref{varprinciple}) is then written
 as an eigenvalue equation,
 $
  \ul \Lm\, \ul c \, b
 = \ul \Gm\, \ul c
 $
 .
\begin{equation}
 \Gm_{pi,qj}
 =
 \int_{S,T}  \psi_{pi}^* \lr{ H - i \pdop t } \psi_{qj}
 + \int_{\prt S, T} \psi_{pi}^* L \psi_{qj}
\end{equation}
\begin{equation}
 \Lm_{pi,qj} =
  \int_{\prt S,T} \psi_{pi}^* \psi_{qj}
\end{equation}
Here $L$ is the usual Bloch operator with the field term,
 defined so that
 $L\Psi = \frac 1r \pder{r \Psi}{r} + \frac ic \mb A \dotp \mh n \Psi$,
 whose eigenvalues are the logarithmic derivative of the reduced wavefunction.
This differs from $\tilde L$ in equation (\ref{tildeLdef})
 only by the addition of one hermitian term,
 so it is still variational.
Only a small subset of the basis functions have nonzero value
 on the surface of the reaction volume.
Denoting these open type basis functions with $o$
 and the others with $c$,
 the eigenvalue equation can be written as follows.
\begin{equation}
 \begin{pmatrix}
  0 & 0 \\ 0 & \ul \Lm_{oo}
 \end{pmatrix}
 \begin{pmatrix}
  \ul c_c \\ \ul c_o
 \end{pmatrix}
 b
 =
 \begin{pmatrix}
  \ul \Gm_{cc} & \ul \Gm_{co} \\ \ul \Gm_{oc} & \ul\Gm_{oo}
 \end{pmatrix}
 \begin{pmatrix}
  \ul c_c \\ \ul c_o
 \end{pmatrix}
\end{equation}
The eigenvalue equation can then be rearranged as follows.
\begin{equation}
 \ul \Lm_{oo} \ul c_o b
 = \ul \Om \, \ul c_o
\qquad
 \ul \Om
 = \ul\Gm_{oo} - \ul\Gm_{oc}\ul\Gm_{cc}^{-1} \ul\Gm_{co}
\end{equation}
This places most of the load of the calculation
 on the linear solution for $\ul\Gm_{cc}^{-1} \ul\Gm_{co} $,
 which requires fewer resources than a full diagonalization,
 although it must be solved for each collision energy of interest.
The matrix $R$ is then calculated from the eigenvalues and eigenvectors
 found in the above.
The resulting $R$ matrix is symmetric,
 and is related to the reduced wavefunction $\ul F(r)$ as follows:
\begin{align} \label{RFW}
 \ul R^{-1} (r) = \ul F'(r) \lr{\ul F(r)}^{-1} + \ul W
 \\
 W_{i,j}
 = \frac ic \int_{\Om, T} \Phi_i^* \mb A\dotp \mh n \Phi_j
\end{align}

\section{Scattering in a laser field} \label{sc_Scattering_laser}

\subsection{Matching to scattering solutions}

The vector potential term
 in the Schr\"odinger equation can be removed
 by the following transformation \cite{Henneberger1968}:
\begin{equation}
 \Phi = \Om \Psi
\qquad
 \Om = \exp\lr{\frac ic \int^t \mb A(\tau) \dotp \mb p 
        \;d\tau}
\end{equation}
After defining
 $\bm \al(t) = \frac 1c \int^t \mb A(\tau) \;d\tau$,
 it becomes clear that $\Om$ is a translation operator,
 i.e. for functions of position,
 $\Om f(\mb r) = f(\mb r + \bm \al(t)) \Om$.
It follows simply that $\Phi$ obeys the equation
\begin{equation}
 i \pder \Phi t
 = - \frac12 \del^2 \Phi + V(\abs{\mb r + \bm \al(t)})
\end{equation}
 which at large $r$ approaches
 the free space Hamiltonian,
 because $V(r) \to 0$ faster than $\frac 1r$.
We may therefore match to free space scattering solutions,
 of the form
\begin{equation}
 \Phi^\bt
 = \sum_{i} Y_{l_i 0} (\tht, \phi)
     \frac{e^{-i(E + \nu_i \om) t}}{\sqrt{2\pi/\om }} 
     \frac{f_i \dl_i^\bt - g_i K_i^\bt}{r}
\end{equation}
Here $\ul K$ is the short range reaction matrix
 and the scattering states are 
\begin{equation} \label{def_scatter_funcs}
 f_i = \sqrt{\frac{2 k_{\nu_i}}{\pi} }
    r\, j_{l_i}\lr{k_{\nu_i} r}
\quad \tand \quad
 g_i = \sqrt{\frac{2 k_{\nu_i}}{\pi} }
    r\, n_{l_i}\lr{k_{\nu_i} r}
\end{equation}
Using the reverse translation,
 $\Psi = \Om^{-1} \Phi$,
 and defining functions $\rho(t)$ and $\tht(t)$
 that describe the length and angle of 
 $\bm{ \rho} (t) = \mb{ r} - \bm{ \al}(t)$
 the velocity gauge wavefunction can be written as
\begin{equation}
 \Psi^\bt
 = \sum_{i} Y_{l_i 0} (\tht(t), \phi)
     \frac{e^{-i(E + \nu_i \om) t}}{\sqrt{2\pi/\om }} 
     \frac{f_i(\rho(t)) - g_i(\rho(t)) K_i^\bt}{\rho(t)}
\end{equation}
Following \citet{VarroEhlotzky1988},
 this can be projected onto the original basis set
 in the untranslated coordinates,
 resulting in
\begin{align}
 \Psi^{\mu \lm}
 &= \sum_{\nu l, \xi j} Y_{l, 0} (\tht, \phi)
     \frac{e^{-i(E + \nu \om) t}}{\sqrt{2\pi/\om }} 
     \frac{\mf{f}_{\nu l}^{\xi j}(r)
	            \dl_{\xi}^\mu \dl_j^\lm
	        - \mf{g}_{\nu l}^{\xi j}(r)
			           K_{\xi j}^{\mu \lm}}{r}
\\
 \label{fdef}
 \mf f_{\nu l}^{\xi j} (r)
  &=
	\sqrt{\frac{2 k_\xi }{\pi}} 
  B_{l , \nu-\xi }^{ \xi  j}
   \,r j_l \lr{k_\xi  r}
\\
 \mf g_{\nu l}^{\xi j} (r)
  &=
	\sqrt{\frac{2 k_\xi }{\pi}} 
  B_{l , \nu-\xi }^{ \xi  j}
   \,r n_l \lr{k_\xi  r}
\\
\nonumber
 B_{ l  s }^{ \xi  j}
 &= \frac{i^{j- l -s}}{2}  \sqrt{\lr{2 j + 1}\lr{2 l  + 1}}
\\ & 
\label{Bdef}
 \hspace{40pt}\int_{-1}^1 dx\, P_j(x) P_ l (x) J_s \lr{-k_\xi  \al_0 x}
\end{align}
The wavefunction can then be expressed
 in matrix form for radial functions,
\begin{equation}
 \ul M(r)
 = \ul{\mf{f}}(r) - \ul{\mf{g}}(r) \ul K
\end{equation}
 and using (\ref{RFW}),
 $\ul K$ can be found in terms of $\ul R$.
\begin{align}
 \ul K = & \lr{\lr{\ul R^{-1}(r_0) - \ul W}\ul{\mf g}(r_0) \
 						- \ul{\mf g'}(r_0) }^{-1}
		\nonumber \\& \hspace{20pt}
 			\lr{\lr{\ul R^{-1}(r_0) - \ul W}\ul{\mf f}(r_0) \
						- \ul R(r_0) \ul{\mf f'}(r_0) }
\end{align}
Note that although $\ul W$, $\ul{\mf f}$, and $\ul{\mf g}$
 are not symmetric,
 this approach yields a reaction matrix that is symmetric and real.

\subsection{Calculating the cross section}

The asymptotic form of the wavefunction in this Floquet picture is
\begin{equation} \label{asymptotic}
 \Psi \lr{\mb r, t}
  = e^{i (k_0 z - E t)}
  + \sum_{\nu \in Z} f_\nu(\tht) \frac{e^{i (k_\nu r - (E + \nu \om)t)}}r
\end{equation}
The cross section for each Floquet channel follows from this.
\begin{equation}
 \der{\sg_\nu}{\Om}
 = \frac{r^2 \abs{\mh r \dotp \mb j_{\tx{out}, \nu}} }{\abs{\mb j_\tx{inc}}}
 = \frac{k_\nu}{k_0} \abs{f_\nu(\tht)}^2
\end{equation}
This can be expressed in terms of the scattering matrix
 $\ul S = \lr{\ul 1 + i \ul K } \lr{\ul 1 - i \ul K}^{-1}$ .
\begin{align}
 &\der{\sg_\nu}{\Om} = \hspace{\fill}
 \nonumber \\
 &\frac{1}{k_0^2}
   \abs{ 
  \sum_{l,l' = 0}^\infty 
   \sqrt{ \pi \lr{2 l' + 1} }
	\lr{
   i^{l' - l} S_{\nu l, 0\, l'}
	- \dl_{l,l'} \dl_{\nu,0}
	}
	 Y_{l,0}(\tht,0) }^2
\end{align}
The phase factor $ i^{l' - l}$ above is a result of the choice
 of phase in the scattering functions (\ref{def_scatter_funcs}).

\section{Results} \label{sc_Results}

\begin{figure} 
\includegraphics[width=3.375in]{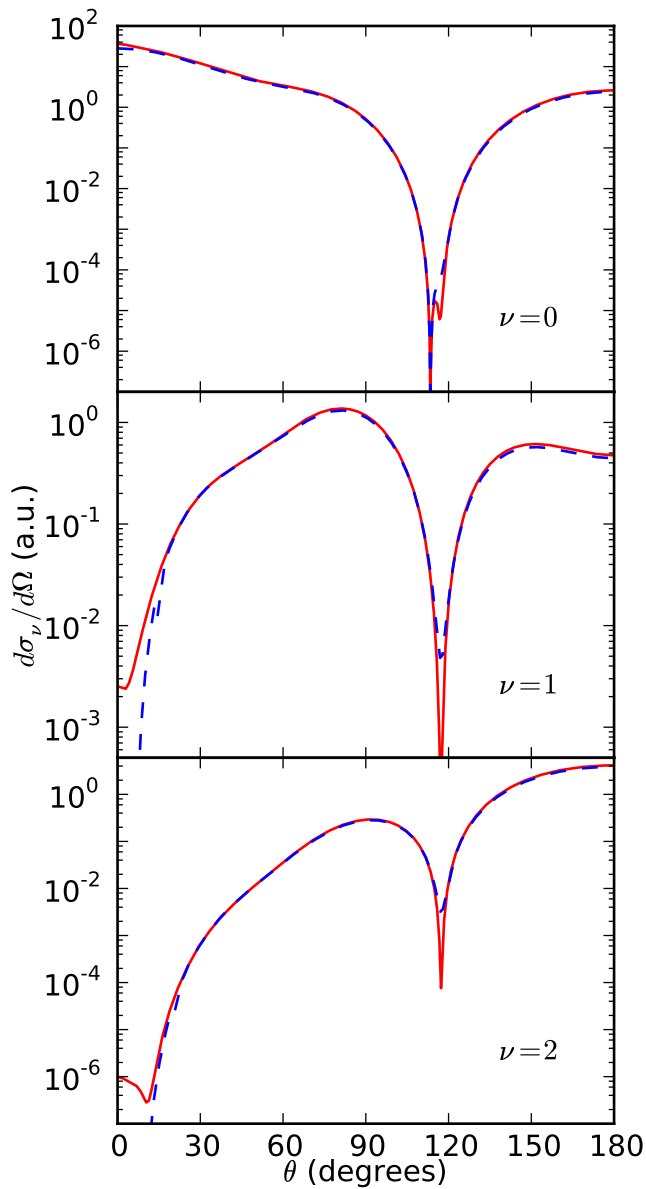}
\caption{ (color online)
 Differential cross section
   for electrons absorbing
   0 (top) to 2 (bottom) photons.
 The solid red line is the cross section
  found with the Floquet $R$ matrix calculation,
  and the dashed blue line is calulated
  using the Kroll Watson approximation.
 This calculation was performed with the $R$ matrix boundary
  at 100 a.u.,
  with 19 total Floquet channels
  and angular momentum channels up to $l = 150$.
      }
\label{xs_KWA}
\end{figure}

Figure \ref{xs_KWA} contains our calculation
 of the cross section
 along with a comparison calculation using
 the Kroll Watson approximation.
The collision parameters were chosen to mimic the experiment:
 a laser intensity of $5\times10^{7} \textrm{ W cm}^{-2}$,
 photon energy of $0.12 \textrm{ eV}$,
 and electron energy $10 \textrm{ eV}$.
It is worth noting that the electron energy lies below
 the first excitation channel of argon,
 so it makes sense to keep just the single channel for the target.
The boundary of the reaction volume for this calculation
 is at $r_0 = 100 \tx{ a.u.}$,
 and it contains 19 Floquet channels up to $\nu = \pm 9$
 and angular momentum channels up to $l = 150$
 for a total of 2869 channels on the surface of the volume.
This number of angular momenta is needed to converge
 the matching equations for the scattering solution;
 as is shown in equation (\ref{fdef}),
 the regular scattering solutions are proportional to $j_l(k r_0)$,
 which has significant value when $l \approx k r_0$.
The number of Floquet channels needed for convergence
 increases with $r_0$ as well,
 but this has more to do with the numerical convergence
 of the $R$ matrix solution than with the matching.
The field is constant in space,
 so a larger reaction volume means more matrix elements
 coupling the Floquet channels.
For Floquet channels $\nu = 1$ and $\nu = 2$,
 the differential cross sections are converged
 with respect to basis size
 to within $10^{-2}$ a.u. and $10^{-3}$ a.u.
 in absolute units of the maximum difference between calculations,
 and to within $1\%$ of the cross section value at all angles.

\begin{figure}
\includegraphics[width=3.375in]{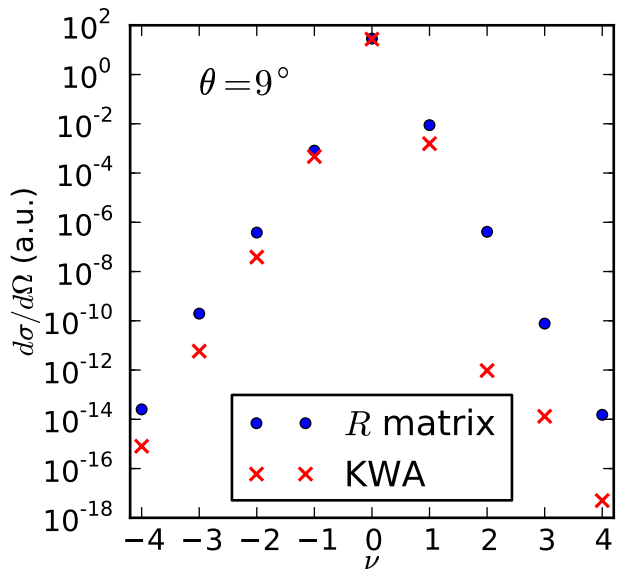}
\caption{(color online)
 Differential cross sections at 9 degrees
  versus the number $\nu$ of photon energies 
  gained by the electron.
 The cross section found via
  the $R$ matrix calculation (blue dots) has comparable
  order of magnitude to the Kroll Watson approximation
  (red crosses) for a few photon numbers,
  and it is several orders of magnitude smaller
  than the experimental results
  for all nonzero $\nu$.
 The parameters of the calculation are the same as for figure \ref{xs_KWA}.
   }
\label{xsd_9}
\end{figure}

Note that the Floquet $R$ matrix calculation and the KWA agree very well
 for all but very small scattering angles.
The zero photon cross section is also quite close
 to the experimental elastic electron argon cross section
 found by \citet{Furst1989}, as expected.
Figure \ref{xsd_9} shows the differential cross section
 at 9 degrees for absorption and emission of up to 4
 photons, again compared with the KWA result.
Note that though they do not agree exactly,
 they differ with far fewer orders of magnitude than
 measured by Wallbank and Holmes at this scattering angle.
For example, the differential cross sections they found
 for exchanges of 1 and 2 photons
 were on the order of 1 percent
 of the field free elastic differential cross section,
 while our calulations show these as roughly 
 $10^{-2}$ percent and $ 10^{-6}$ percent respectively.

Differential cross sections calculated with $R$ matrix boundaries
 from 10 a.u.\ to 100 a.u.\ show no differences
 that are distinguishable from the convergence with respect to the basis.
Our differential cross section also agrees with
 that of \citet{ChenRobicheaux96},
 who used a variable gauge approach
 and chose an $R$ matrix boundary of 30 a.u.
We can, therefore, rule out the possibility
 that the long range induced-dipole potential
 would yield the sort of diffraction
 suggested by \citet{Jaron1999}
 over distances comparable to or even several times
 the electron de Broglie wavelength.

It has been suggested 
 that uncertainty in the scattering angle could account for
 higher observed cross sections.
Madsen \cite{Madsen1995} even suggests that
 the incoming electron beam is poorly collimated,
 leading to an effective error in the scattering angle as high as $8^\circ$,
 as opposed to the $2^\circ$
 reported from the detector width \cite{Wallbank1994Ar}.
Convolving the cross sections shown in figure \ref{xs_KWA}
 with a gaussian having a width up to $8^\circ$
 does not give a significant difference in the
 1- and 2-photon cross sections, however.
Whatever the source of such an error,
 uncertanties in the scattering angle would not explain
 the experimental cross sections for this geometry.

\citet{Sun1998} calculate a cross section for one photon exchange
 that is quantitatively quite close to our result.
Their result for two photon exchange is several orders of magnitude higher,
 however.
This may be due to a convergence issue,
 as a group using a similar method for laser-assisted helium scattering
 at first found high cross sections \cite{Jaron1997},
 but later found better results that are closer to the KWA \cite{Garland2002}.

\section{Conclusions} \label{sc_Conclusions}

The Floquet $R$ matrix method provides an exact solution
 of the Scr\"odinger equation in the velocity gauge.
By comparing cross sections calculated with $R$ matrix boundaries
 up to 100 a.u.,
 over ten times the de Broglie wavelength of a 10 eV electron,
 we have shown that the induced dipole potential for argon
 does not contribute to the laser assisted cross section.
This is true even at small angles
 where the momentum transfer has a very small component along the field,
 and so the KWA is less valid.
Diffraction from this long range piece of the potential
 can not account for the high cross sections
 found in the experiments of Wallbank and Holmes,
 which are several orders of magnitude higher
 than both the approximation and our calculations.

The most plausible explanation for the experimental results
 remains multiple scattering.
Later experiments for helium by the same group claim
 to see the same high cross sections
 even when the gas is too dilute for multiple scattering
 to play a significant role \cite{Wallbank2001},
 so it is unclear whether this is the correct explanation.
To our knowledge, a similar experiment for argon
 including characterization of the gas density has not been performed.
 
Discussions with A. Jaron-Becker and M. Tarana
 are much appreciated.
We gratefully acknowledge the support
 of the Department of Energy,
 Office of Science.


\begin{thebibliography}{19}%
\makeatletter
\providecommand \@ifxundefined [1]{%
 \@ifx{#1\undefined}
}%
\providecommand \@ifnum [1]{%
 \ifnum #1\expandafter \@firstoftwo
 \else \expandafter \@secondoftwo
 \fi
}%
\providecommand \@ifx [1]{%
 \ifx #1\expandafter \@firstoftwo
 \else \expandafter \@secondoftwo
 \fi
}%
\providecommand \natexlab [1]{#1}%
\providecommand \enquote  [1]{``#1''}%
\providecommand \bibnamefont  [1]{#1}%
\providecommand \bibfnamefont [1]{#1}%
\providecommand \citenamefont [1]{#1}%
\providecommand \href@noop [0]{\@secondoftwo}%
\providecommand \href [0]{\begingroup \@sanitize@url \@href}%
\providecommand \@href[1]{\@@startlink{#1}\@@href}%
\providecommand \@@href[1]{\endgroup#1\@@endlink}%
\providecommand \@sanitize@url [0]{\catcode `\\12\catcode `\$12\catcode
  `\&12\catcode `\#12\catcode `\^12\catcode `\_12\catcode `\%12\relax}%
\providecommand \@@startlink[1]{}%
\providecommand \@@endlink[0]{}%
\providecommand \url  [0]{\begingroup\@sanitize@url \@url }%
\providecommand \@url [1]{\endgroup\@href {#1}{\urlprefix }}%
\providecommand \urlprefix  [0]{URL }%
\providecommand \Eprint [0]{\href }%
\providecommand \doibase [0]{http://dx.doi.org/}%
\providecommand \selectlanguage [0]{\@gobble}%
\providecommand \bibinfo  [0]{\@secondoftwo}%
\providecommand \bibfield  [0]{\@secondoftwo}%
\providecommand \translation [1]{[#1]}%
\providecommand \BibitemOpen [0]{}%
\providecommand \bibitemStop [0]{}%
\providecommand \bibitemNoStop [0]{.\EOS\space}%
\providecommand \EOS [0]{\spacefactor3000\relax}%
\providecommand \BibitemShut  [1]{\csname bibitem#1\endcsname}%
\let\auto@bib@innerbib\@empty
\bibitem [{\citenamefont {Kroll}\ and\ \citenamefont
  {Watson}(1973)}]{KrollWatson1973}%
  \BibitemOpen
  \bibfield  {author} {\bibinfo {author} {\bibfnamefont {N.~M.}\ \bibnamefont
  {Kroll}}\ and\ \bibinfo {author} {\bibfnamefont {K.~M.}\ \bibnamefont
  {Watson}},\ }\href {\doibase 10.1103/PhysRevA.8.804} {\bibfield  {journal}
  {\bibinfo  {journal} {Phys. Rev. A}\ }\textbf {\bibinfo {volume} {8}},\
  \bibinfo {pages} {804} (\bibinfo {year} {1973})}\BibitemShut {NoStop}%
\bibitem [{\citenamefont {Wallbank}\ and\ \citenamefont
  {Holmes}(1993)}]{Wallbank1993}%
  \BibitemOpen
  \bibfield  {author} {\bibinfo {author} {\bibfnamefont {B.}~\bibnamefont
  {Wallbank}}\ and\ \bibinfo {author} {\bibfnamefont {J.~K.}\ \bibnamefont
  {Holmes}},\ }\href {\doibase 10.1103/PhysRevA.48.R2515} {\bibfield  {journal}
  {\bibinfo  {journal} {Phys. Rev. A}\ }\textbf {\bibinfo {volume} {48}},\
  \bibinfo {pages} {R2515} (\bibinfo {year} {1993})}\BibitemShut {NoStop}%
\bibitem [{\citenamefont {Wallbank}\ and\ \citenamefont
  {Holmes}(1994{\natexlab{a}})}]{Wallbank1994}%
  \BibitemOpen
  \bibfield  {author} {\bibinfo {author} {\bibfnamefont {B.}~\bibnamefont
  {Wallbank}}\ and\ \bibinfo {author} {\bibfnamefont {J.~K.}\ \bibnamefont
  {Holmes}},\ }\href {http://stacks.iop.org/0953-4075/27/i=6/a=020} {\bibfield
  {journal} {\bibinfo  {journal} {J. Phys. B}\ }\textbf {\bibinfo {volume}
  {27}},\ \bibinfo {pages} {1221} (\bibinfo {year}
  {1994}{\natexlab{a}})}\BibitemShut {NoStop}%
\bibitem [{\citenamefont {Wallbank}\ and\ \citenamefont
  {Holmes}(1994{\natexlab{b}})}]{Wallbank1994Ar}%
  \BibitemOpen
  \bibfield  {author} {\bibinfo {author} {\bibfnamefont {B.}~\bibnamefont
  {Wallbank}}\ and\ \bibinfo {author} {\bibfnamefont {J.~K.}\ \bibnamefont
  {Holmes}},\ }\href {http://stacks.iop.org/0953-4075/27/i=21/a=027} {\bibfield
   {journal} {\bibinfo  {journal} {J. Phys. B}\ }\textbf {\bibinfo {volume}
  {27}},\ \bibinfo {pages} {5405} (\bibinfo {year}
  {1994}{\natexlab{b}})}\BibitemShut {NoStop}%
\bibitem [{\citenamefont {Rabad\'an}\ \emph {et~al.}(1994)\citenamefont
  {Rabad\'an}, \citenamefont {M\'endez},\ and\ \citenamefont
  {Dickinson}}]{Rabadan1994}%
  \BibitemOpen
  \bibfield  {author} {\bibinfo {author} {\bibfnamefont {I.}~\bibnamefont
  {Rabad\'an}}, \bibinfo {author} {\bibfnamefont {L.}~\bibnamefont {M\'endez}},
  \ and\ \bibinfo {author} {\bibfnamefont {A.~S.}\ \bibnamefont {Dickinson}},\
  }\href {http://stacks.iop.org/0953-4075/27/i=16/a=006} {\bibfield  {journal}
  {\bibinfo  {journal} {J. Phys. B}\ }\textbf {\bibinfo {volume} {27}},\
  \bibinfo {pages} {L535} (\bibinfo {year} {1994})}\BibitemShut {NoStop}%
\bibitem [{\citenamefont {Geltman}(1995)}]{Geltman1995}%
  \BibitemOpen
  \bibfield  {author} {\bibinfo {author} {\bibfnamefont {S.}~\bibnamefont
  {Geltman}},\ }\href {\doibase 10.1103/PhysRevA.51.R34} {\bibfield  {journal}
  {\bibinfo  {journal} {Phys. Rev. A}\ }\textbf {\bibinfo {volume} {51}},\
  \bibinfo {pages} {R34} (\bibinfo {year} {1995})}\BibitemShut {NoStop}%
\bibitem [{\citenamefont {Varr\'o}\ and\ \citenamefont
  {Ehlotzky}(1995)}]{Varro1995}%
  \BibitemOpen
  \bibfield  {author} {\bibinfo {author} {\bibfnamefont {S.}~\bibnamefont
  {Varr\'o}}\ and\ \bibinfo {author} {\bibfnamefont {F.}~\bibnamefont
  {Ehlotzky}},\ }\href {\doibase 10.1016/0375-9601(95)00357-9} {\bibfield
  {journal} {\bibinfo  {journal} {Phys. Lett.}\ }\textbf {\bibinfo {volume}
  {203}},\ \bibinfo {pages} {203 } (\bibinfo {year} {1995})}\BibitemShut
  {NoStop}%
\bibitem [{\citenamefont {Rabad\'an}\ \emph {et~al.}(1996)\citenamefont
  {Rabad\'an}, \citenamefont {M\'endez},\ and\ \citenamefont
  {Dickinson}}]{Dickinson1996dbl}%
  \BibitemOpen
  \bibfield  {author} {\bibinfo {author} {\bibfnamefont {I.}~\bibnamefont
  {Rabad\'an}}, \bibinfo {author} {\bibfnamefont {L.}~\bibnamefont {M\'endez}},
  \ and\ \bibinfo {author} {\bibfnamefont {A.~S.}\ \bibnamefont {Dickinson}},\
  }\href {http://stacks.iop.org/0953-4075/29/i=22/a=004} {\bibfield  {journal}
  {\bibinfo  {journal} {J. Phys. B}\ }\textbf {\bibinfo {volume} {29}},\
  \bibinfo {pages} {L801} (\bibinfo {year} {1996})}\BibitemShut {NoStop}%
\bibitem [{\citenamefont {Madsen}\ and\ \citenamefont
  {Taulbjerg}(1995)}]{Madsen1995}%
  \BibitemOpen
  \bibfield  {author} {\bibinfo {author} {\bibfnamefont {L.~B.}\ \bibnamefont
  {Madsen}}\ and\ \bibinfo {author} {\bibfnamefont {K.}~\bibnamefont
  {Taulbjerg}},\ }\href {http://stacks.iop.org/0953-4075/28/i=24/a=017}
  {\bibfield  {journal} {\bibinfo  {journal} {J. Phys. B}\ }\textbf {\bibinfo
  {volume} {28}},\ \bibinfo {pages} {5327} (\bibinfo {year}
  {1995})}\BibitemShut {NoStop}%
\bibitem [{\citenamefont {Sun}\ \emph {et~al.}(1998)\citenamefont {Sun},
  \citenamefont {Zhang}, \citenamefont {Jiang},\ and\ \citenamefont
  {Yu}}]{Sun1998}%
  \BibitemOpen
  \bibfield  {author} {\bibinfo {author} {\bibfnamefont {J.}~\bibnamefont
  {Sun}}, \bibinfo {author} {\bibfnamefont {S.}~\bibnamefont {Zhang}}, \bibinfo
  {author} {\bibfnamefont {Y.}~\bibnamefont {Jiang}}, \ and\ \bibinfo {author}
  {\bibfnamefont {G.}~\bibnamefont {Yu}},\ }\href {\doibase
  10.1103/PhysRevA.58.2225} {\bibfield  {journal} {\bibinfo  {journal} {Phys.
  Rev. A}\ }\textbf {\bibinfo {volume} {58}},\ \bibinfo {pages} {2225}
  (\bibinfo {year} {1998})}\BibitemShut {NoStop}%
\bibitem [{\citenamefont {Jaro\'n}\ and\ \citenamefont
  {Kami\'nski}(1997)}]{Jaron1997}%
  \BibitemOpen
  \bibfield  {author} {\bibinfo {author} {\bibfnamefont {A.}~\bibnamefont
  {Jaro\'n}}\ and\ \bibinfo {author} {\bibfnamefont {J.~Z.}\ \bibnamefont
  {Kami\'nski}},\ }\href {\doibase 10.1103/PhysRevA.56.R4393} {\bibfield
  {journal} {\bibinfo  {journal} {Phys. Rev. A}\ }\textbf {\bibinfo {volume}
  {56}},\ \bibinfo {pages} {R4393} (\bibinfo {year} {1997})}\BibitemShut
  {NoStop}%
\bibitem [{\citenamefont {Jaro\'n}\ and\ \citenamefont
  {Kami\'nski}(1999)}]{Jaron1999}%
  \BibitemOpen
  \bibfield  {author} {\bibinfo {author} {\bibfnamefont {A.}~\bibnamefont
  {Jaro\'n}}\ and\ \bibinfo {author} {\bibfnamefont {J.~Z.}\ \bibnamefont
  {Kami\'nski}},\ }\href@noop {} {\bibfield  {journal} {\bibinfo  {journal}
  {Laser Physics}\ }\textbf {\bibinfo {volume} {9}},\ \bibinfo {pages} {81}
  (\bibinfo {year} {1999})}\BibitemShut {NoStop}%
\bibitem [{\citenamefont {Chen}\ and\ \citenamefont
  {Robicheaux}(1996)}]{ChenRobicheaux96}%
  \BibitemOpen
  \bibfield  {author} {\bibinfo {author} {\bibfnamefont {C.-T.}\ \bibnamefont
  {Chen}}\ and\ \bibinfo {author} {\bibfnamefont {F.}~\bibnamefont
  {Robicheaux}},\ }\href {http://stacks.iop.org/0953-4075/29/i=2/a=022}
  {\bibfield  {journal} {\bibinfo  {journal} {J. Phys. B}\ }\textbf {\bibinfo
  {volume} {29}},\ \bibinfo {pages} {345} (\bibinfo {year} {1996})}\BibitemShut
  {NoStop}%
\bibitem [{\citenamefont {Tolstikhin}\ \emph {et~al.}(1996)\citenamefont
  {Tolstikhin}, \citenamefont {Watanabe},\ and\ \citenamefont
  {Matsuzawa}}]{Tolstikhin1996}%
  \BibitemOpen
  \bibfield  {author} {\bibinfo {author} {\bibfnamefont {O.~I.}\ \bibnamefont
  {Tolstikhin}}, \bibinfo {author} {\bibfnamefont {S.}~\bibnamefont
  {Watanabe}}, \ and\ \bibinfo {author} {\bibfnamefont {M.}~\bibnamefont
  {Matsuzawa}},\ }\href {http://stacks.iop.org/0953-4075/29/i=11/a=001}
  {\bibfield  {journal} {\bibinfo  {journal} {J. Phys. B}\ }\textbf {\bibinfo
  {volume} {29}},\ \bibinfo {pages} {L389} (\bibinfo {year}
  {1996})}\BibitemShut {NoStop}%
\bibitem [{\citenamefont {Henneberger}(1968)}]{Henneberger1968}%
  \BibitemOpen
  \bibfield  {author} {\bibinfo {author} {\bibfnamefont {W.~C.}\ \bibnamefont
  {Henneberger}},\ }\href {\doibase 10.1103/PhysRevLett.21.838} {\bibfield
  {journal} {\bibinfo  {journal} {Phys. Rev. Lett.}\ }\textbf {\bibinfo
  {volume} {21}},\ \bibinfo {pages} {838} (\bibinfo {year} {1968})}\BibitemShut
  {NoStop}%
\bibitem [{\citenamefont {Varr\'o}\ and\ \citenamefont
  {Ehlotzky}(1988)}]{VarroEhlotzky1988}%
  \BibitemOpen
  \bibfield  {author} {\bibinfo {author} {\bibfnamefont {S.}~\bibnamefont
  {Varr\'o}}\ and\ \bibinfo {author} {\bibfnamefont {F.}~\bibnamefont
  {Ehlotzky}},\ }\href {http://dx.doi.org/10.1007/BF01436943} {\bibfield
  {journal} {\bibinfo  {journal} {Z. Phys. D}\ }\textbf {\bibinfo {volume}
  {8}},\ \bibinfo {pages} {211} (\bibinfo {year} {1988})}\BibitemShut {NoStop}%
\bibitem [{\citenamefont {Furst}\ \emph {et~al.}(1989)\citenamefont {Furst},
  \citenamefont {Golden}, \citenamefont {Mahgerefteh}, \citenamefont {Zhou},\
  and\ \citenamefont {Mueller}}]{Furst1989}%
  \BibitemOpen
  \bibfield  {author} {\bibinfo {author} {\bibfnamefont {J.~E.}\ \bibnamefont
  {Furst}}, \bibinfo {author} {\bibfnamefont {D.~E.}\ \bibnamefont {Golden}},
  \bibinfo {author} {\bibfnamefont {M.}~\bibnamefont {Mahgerefteh}}, \bibinfo
  {author} {\bibfnamefont {J.}~\bibnamefont {Zhou}}, \ and\ \bibinfo {author}
  {\bibfnamefont {D.}~\bibnamefont {Mueller}},\ }\href {\doibase
  10.1103/PhysRevA.40.5592} {\bibfield  {journal} {\bibinfo  {journal} {Phys.
  Rev. A}\ }\textbf {\bibinfo {volume} {40}},\ \bibinfo {pages} {5592}
  (\bibinfo {year} {1989})}\BibitemShut {NoStop}%
\bibitem [{\citenamefont {Garland}\ \emph {et~al.}(2002)\citenamefont
  {Garland}, \citenamefont {Jaron}, \citenamefont {Kaminski},\ and\
  \citenamefont {Potvliege}}]{Garland2002}%
  \BibitemOpen
  \bibfield  {author} {\bibinfo {author} {\bibfnamefont {L.~W.}\ \bibnamefont
  {Garland}}, \bibinfo {author} {\bibfnamefont {A.}~\bibnamefont {Jaron}},
  \bibinfo {author} {\bibfnamefont {J.~Z.}\ \bibnamefont {Kaminski}}, \ and\
  \bibinfo {author} {\bibfnamefont {R.~M.}\ \bibnamefont {Potvliege}},\ }\href
  {http://stacks.iop.org/0953-4075/35/i=13/a=302} {\bibfield  {journal}
  {\bibinfo  {journal} {J. Phys. B}\ }\textbf {\bibinfo {volume} {35}},\
  \bibinfo {pages} {2861} (\bibinfo {year} {2002})}\BibitemShut {NoStop}%
\bibitem [{\citenamefont {Wallbank}\ and\ \citenamefont
  {Holmes}(2001)}]{Wallbank2001}%
  \BibitemOpen
  \bibfield  {author} {\bibinfo {author} {\bibfnamefont {B.}~\bibnamefont
  {Wallbank}}\ and\ \bibinfo {author} {\bibfnamefont {J.~K.}\ \bibnamefont
  {Holmes}},\ }\href
  {http://www.ingentaconnect.com/content/nrc/cjp/2001/00000079/00000010/art00005}
  {\bibfield  {journal} {\bibinfo  {journal} {Can. J. Phys.}\ }\textbf
  {\bibinfo {volume} {79}},\ \bibinfo {pages} {1237} (\bibinfo {year}
  {2001})}\BibitemShut {NoStop}%
\end{thebibliography}
\end{document}